\documentclass[aps,twocolumn,showpacs]{revtex4}
\usepackage{graphicx}
\usepackage{dcolumn}
\usepackage{bm}

\begin{document}

\title{Phase Diagram And Adsorption-Desorption Kinetics Of $CO$ On $Ru(0001)$: \\Present Limitations Of A First Principles Approach}

\author{J.-S. McEwen}
      \email{mcewenj@fizz.phys.dal.ca}
 \affiliation{Department of Physics and Atmospheric Science,
       Dalhousie University,
       Halifax, Nova Scotia,
       Canada, B3H 3J5}

\author{A. Eichler}
\affiliation{Institut f\"{u}r Materialphysik and Centre for Computational Materials Science,
Universit\"{a}t Wien,
Sensengasse 8/12 A-1090 Wien, Austria
}
\date{\today}

\begin{abstract}
A lattice gas model is used to study the equilibrium properties and desorption
kinetics of CO on Ru(0001). With interactions obtained from
density functional theory (DFT) the phase diagram and temperature
programmed desorption (TPD) spectra are calculated up to a
coverage of 1/3 ML using top sites only. For coverages beyond 1/3
ML hollow sites are included. Good agreement is obtained between
experiment and theory for coverages below 1/3 ML using top sites
only. When including hollow sites, DFT calculations fail in
predicting the correct binding energy differences between top and
hollow sites giving disagreement with TPD, low energy electron
diffraction (LEED) and heat of adsorption experiments.
\end{abstract}
\maketitle

\section{Introduction}
     Adsorbate systems with commensurate structures can be described successfully
     with a lattice gas model with one or more adsorption sites. The energetics are
     contained in a few parameters such as binding energies, vibrational frequencies
     and lateral interactions. In a ``complete'' theory these parameters would be
     obtained from first principles quantum mechanical calculations, for instance
     based on density functional theory. This has been demonstrated successfully
     for oxygen on Ru(0001) for both equilibrium and desorption kinetics
     \cite{cathy,mcewen}. An attempt to repeat such an approach for CO/Pt(111)
     has been shown to fail \cite{mcewen2}, most importantly due to the fact that
     DFT does not produce the right
     energetic order of the binding sites for many transition and noble metal substrates
     \cite{Feibelmann01,marek,kresse}. Nevertheless a satisfactory explanation of all available
     equilibrium and kinetic data could be achieved by treating a minimal set of the
     lattice gas parameters as phenomenological fitting parameters \cite{mcewen2}.
     In this paper we will show that for CO/Ru(0001) DFT
     yields parameters that give a good account for data below 1/3 ML
     provided only top sites are included, but fails when bridge and hollow sites
     come into play. \\ \\
     Let us begin with a quick survey of the experimental data needed
     to model this system appropriately; details and actual data will be given later.
     At low coverage ($\theta \le 1/3$ ML) and temperature
     (T $<$ 150 K)  CO binds to on-top adsorption sites only resulting in a
     $(\sqrt{3}\times\sqrt{3})R30^{\circ}$ ordered
     structure at 1/3 ML \cite{pfnur,menzel,weinberg,menzel2}.
     As the coverage is increased beyond
     1/3 ML three fold hollow sites become occupied forming a
     $(2\sqrt{3}\times2\sqrt{3})R30^{\circ}$ structure at 1/2 ML, with equal population
     of top, hcp and fcc sites, for which
     two different geometries have been proposed \cite{steinruck,pfnur2}.
     For coverages beyond 1/2 ML, a p($7\times7$) structure at 0.55 ML
     with top and bridge sites \cite{steinruck} and a $(2\sqrt{3}\times2\sqrt{3})R30^{\circ}$
     structure at 7/12 ML \cite{steinruck,pfnur2} have been
     suggested.  Finally, a $(5\sqrt{3}\times5\sqrt{3})R30^{\circ}$ saturation
     structure at 49/75 ML was observed
     using He scattering experiments \cite{Woll}.  A summary phase
     diagram was given by Pfn\"{u}r {\it et al.} \cite{pfnur2}.  Equilibrium
     isobars and the isosteric heat of adsorption \cite{menzel3} have been measured,
     too, as well as the temperature programmed
     desorption spectra \cite{menzel3,steinruck2}.  The sticking coefficient
     was measured directly with molecular beam scattering
     as a function of temperature, coverage and kinetic energy of the
     incident beam \cite{steinruck2} and by coverage vs. exposure curves
     \cite{menzelstick}.  Measurements have also been done
     for the CO stretch
     frequency using infrared reflection-adsorption
     spectroscopy (IRAS) \cite{menzel2,weinberg2,jacob2}.  The other frequency modes of the
     on top species have been determined
     using helium scattering experiments \cite{Woll} and IRAS measurements \cite{jakob}.

     In the next section we briefly introduce the lattice gas model, the DFT setup as well
     as the methods used to calculate the equilibrium and kinetic data.  In section 3 we
     consider the modeling of CO/Ru(0001) with top sites only up to 1/3 ML
     and in section 4 we show and discuss the modeling of this
     system with top and hollow binding sites up to 1/2 ML.  The paper ends with a
     discussion to what extent present DFT methods can be used to model CO/Ru(0001).
     \section{Theoretical Methods}
     \subsection{Lattice gas formalism}
     To set up a lattice gas model for CO/Ru(0001) we require a Hamiltonian, which we write
     down for top sites only as:
     \begin{eqnarray}
       H &=& -\left(V_0+k_BT\ln(q_{\rm 3}q_{\rm int})\right) \sum_i t_i \\ \nonumber
        &+& \frac{1}{2}\sum_n^3 \sum_i \sum_{a_n} V_n t_i t_{i+a_{n}}
       + \frac{1}{3}\sum_i \sum_{a,b} V_{\rm trio} t_i t_{i+a} t_{i+b}
     \end{eqnarray}
     with obvious generalizations to a multi-site system \cite{cathy,mcewen2}.
     Here the sum on $i$ exhausts all lattice cells;
     $i+a_n$, $i+a$ and $i+b$ label neighboring cells and we have introduced occupation
     numbers $t_i=0$ or 1 depending on whether a site in cell $i$ is empty or occupied.
$V_0$ represents the
     depth of the adsorption potential well with reference to the gas phase molecule.     Moreover,
     $V_{1}$, $V_{2}$ and $V_{3}$
     denote first, second and third nearest neighbor interactions respectively.
     $V_{\rm trio}$ includes $V_{lt}$, $V_{bt}$,
     $V_{tt}$ which denote linear, bent and triangular trio interactions involving first and
     sometimes second nearest neighbor interactions \cite{cathy,mcewen2}.
     Furthermore, $q_{\rm 3}=q_{\rm z}q_{\rm xy}$ is the vibrational partition function
     of the adsorbed molecule for its center of mass vibrations with respect to
     the surface with $q_{\rm z}$ being the  component perpendicular to it.  Likewise,
     $q_{\rm xy}$ is the partition function for the motion parallel to the surface.  We
     have also allowed for the fact that the internal partition for rotations and
     vibrations of an adsorbed molecule is changed from its free gas phase value,
     $Z_{\rm int}$ to  $q_{\rm int}$, if some of the internal degrees of freedom are
     frozen out or frustrated \cite{kreuzer_review}.\\ \\
     We determined the temperature-coverage phase diagram by calculating the corresponding
      $(\sqrt{3}\times\sqrt{3})R30^{\circ}$ order parameter \cite{landau}.
      Second order
     phase transitions were defined at a given temperature by the inflection point of the
     order parameter \cite{piercy,mouritsen} as a function of the chemical potential.
     First order phase transitions were marked
     when it was discontinuous \cite{mouritsen,landau2}.
     To calculate the resulting TPD spectra we used
     the theoretical framework described in
     \cite{kreuzer_review}.  The desorption rate depending essentially on the sticking
     coefficient $S(\theta, T)$ and the chemical potential of the adsorbate
     $\mu(\theta, T)$ (which includes its binding energy and the frequencies
     of the CO molecule
     in the gas phase and on the surface).
     \subsection{Monte Carlo Methods}
The TPD, isobars and the phase diagram were calculated with Monte Carlo
methods. Simulations were performed in
both the grand canonical and canonical ensembles (chemical
potential and coverage specified, respectively) using the
Metropolis algorithm (with spin flip and infinite Kawasaki
dynamics) \cite{binder}. Equilibration times of the order $2^{14}$
up to $2^{18}$ Monte Carlo sweeps were allowed for each coverage or chemical
potential point. When dealing with top sites only, we initialized
the system in a $(\sqrt{3}\times\sqrt{3})R30^{\circ}$ structure
for the first temperature-coverage, temperature-chemical potential
point, respectively.  For coverages beyond 1/3 ML, we occasionally
initialized the system with a p$(2 \times 2)$
structure (with equal populations of top and hcp sites at 1/2 ML)
or with a clean substrate. Thereafter, to fasten equilibration,
initialization of the system occurred with the final configuration
of the previously calculated point. When calculating the phase
diagram, we performed averages over a maximum of five independent
samples. For the canonical ensemble, the particle insertion method
of Widom \cite{widom} was used in order to calculate the chemical
potential with Monte Carlo methods; a method which was developed
to be applicable especially for the interaction
     parameters derived from DFT calculations.
To allow adsorption on hollow and top sites we used three
interpenetrating $42\times 42$ or $60\times 60$ lattices, with periodic boundary
conditions. This lattice size was chosen to allow commensurate
low-coverage ordered structures with periods of 3 and 6.

\subsection{Density Functional Theory setup}
     We have calculated all necessary interaction parameters with the Vienna ab initio
     simulation package VASP \cite{kre96a,kre96b,vasp}, a plane-wave DFT program,
    which is based on the projector augmented wave
     method \cite{kre98}.
 A cut-off energy for the expansion of
the plane waves of 400 eV was found to be sufficient for an
accurate description. For exchange and correlation the generalized
gradient approximation (GGA) according to Perdew et al.
\cite{per92} was applied. The surface was modeled in $p(2\times2)$
and $p(3\times3)$ cells at the theoretical lattice constant of
a=2.725~\AA~and c/a=1.579, with slabs consisting out of 6 layers,
of which the uppermost two layers were structurally optimized
where mentioned in the text. The Brillouin zone was sampled by a
grid of $(3\times2\times1)$ k-points.

 The adsorption energy for an isolated
     CO molecule ($-V_0$) is defined as the adsorption energy for 1/9 ML coverage ($E_a^{\theta=1/9}$)
derived from the total energies of the adsorbate/substrate system
($E_{total}^{CO/Ru}$), the bare surface ($E_{total}^{Ru}$) and the
free CO molecule ($E_{total}^{CO}$):
     \begin{equation}
        -V_0 = E_a^{\theta=1/9} = E_{total}^{CO/Ru} - E_{total}^{Ru} -
       E_{total}^{CO}
     \end{equation}
     The interaction parameters where determined by calculating the energy of the
     system for ten configurations at various coverages between 1/9 and
     1 ML \cite{cathy}.  The
     energy of the system at a coverage of 1 ML was
     calculated twice: once with the $(3\times3)$ unit cell and again with a
     $(2\times2)$ unit cell.  Consequently, all our
     interaction sets had more ordered structures than lattice
     gas Hamiltonian parameters so that we have fitted, using a least squares procedure,
     for the resulting interactions.  Maximum deviations from the calculated energies
     of the ordered structures with respect to what was obtained from the
     interaction parameters, were never greater than 5 meV.  \\ \\
\section{Modeling with on-top sites at low coverage}
     CO/Ru(0001) has already been modeled successfully in the past using a lattice
      gas model with top sites only.  The interaction and binding energy parameters
      were obtained phenomenologically by matching the
      phase diagram \cite{nagai}, the TPD spectra and
     the isosteric heat of
     adsorption \cite{kreuzer5}.

     In this work, three strategies were used to obtain these
     energies:
\begin{itemize}
\item In the first ({\em method A}) the top two layers
     of the bare substrate were completely relaxed and then kept fixed at these positions
     for the description of CO adsorption.
     The geometry of the adsorbed CO molecule was optimized only for the 1/9 ML
     configuration.    For higher
     coverages, the same CO bond length and adsorption height was
     used.
\item In {\em method B} the position of the CO molecules as well
as the first two layers of the substrate were allowed to relax for
all of the ordered structures. \item     For comparison, we have
also looked
     at the resulting interaction parameters when fixing laterally the CO molecules
     for the structure at 2/9 ML (which is the only configuration for which
     lateral forces on the CO molecule are symmetrically allowed) with all the other ordered structures
     fully relaxed (the
     CO molecules and the first two layers of the substrate).  This third approach ({\em method C}) was
     used for O/Ru(0001) \cite{cathy}.\end{itemize}

     In Table 1, we show the resulting energies
     using all of the above three
     methods together with the standard deviation $\sigma$ describing the average
     deviation between the calculated
     energies and the parametrization.  We used the minimum
     number of parameters so as to have an acceptable value of $\sigma$.
     These
     interaction sets show that the second neighbor interaction between top
     sites, $V_2$, is repulsive for the relaxed calculations.  This is clearly at odds
     with the experimental phase diagram which implies a coexistence between the
     $(\sqrt{3}\times\sqrt{3})R30^{\circ}$ structure and a lattice gas, and
     requires $V_2$ attractive.  In addition, the
     calculated TPD spectra using transfer matrix techniques
     with method B and C give low initial-coverage spectra peak
     positions that do not correspond to experiment and are much too broad.
     On a more positive note,
     the first neighbor interaction between ontop sites for method B (cf. Table
     \ref{tab;intdft2})
     is closer to what
     was obtained by previous phenomenological models of this system
     using only one binding site \cite{kreuzer5}.

Interestingly, for method C $V_2$ becomes negative ($-$6 meV) if
the same set of parameters ($V_1$, $V_2$, $V_{lt}$, $V_{tt}$) is
used as for method A. However, in this case the standard deviation
increases significantly to 19 meV.

Because of this obvious discrepancy for the relaxed calculations,
we restrict the further analysis, to the results obtained within
method A. The reason for this behavior can be found probably in
local relaxations for specific configurations used for the
determination of the parameters, which are not representative for
the interaction of CO on Ru(0001) in general. For the unrelaxed
setup in method A, the CO molecule and the surface is treated
completely identically in all configurations, entirely in the
spirit of the lattice gas like interaction.  For the relaxed
setups, however, the number of input configurations would probably
have to be increased such that relaxation effects, which are only
characteristic of one specific arrangements of molecules, but not
of the interaction between molecules itself, are averaged out.

\begin{figure} \begin{center}
    \includegraphics*[bb=48 267 464 695, width = 5.5 cm]{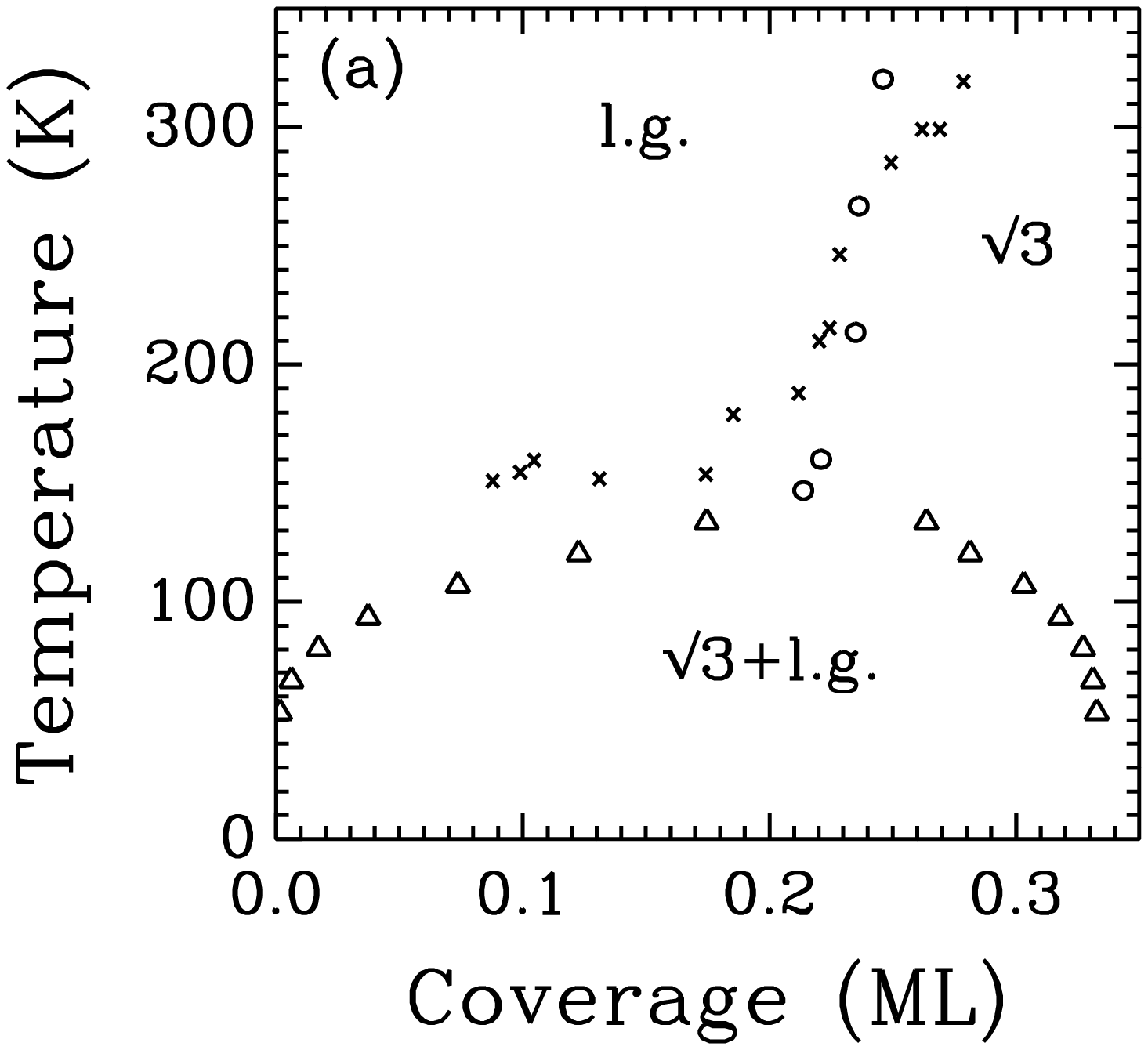}\\
    \includegraphics*[bb=48 267 464 695, width = 5.5 cm]{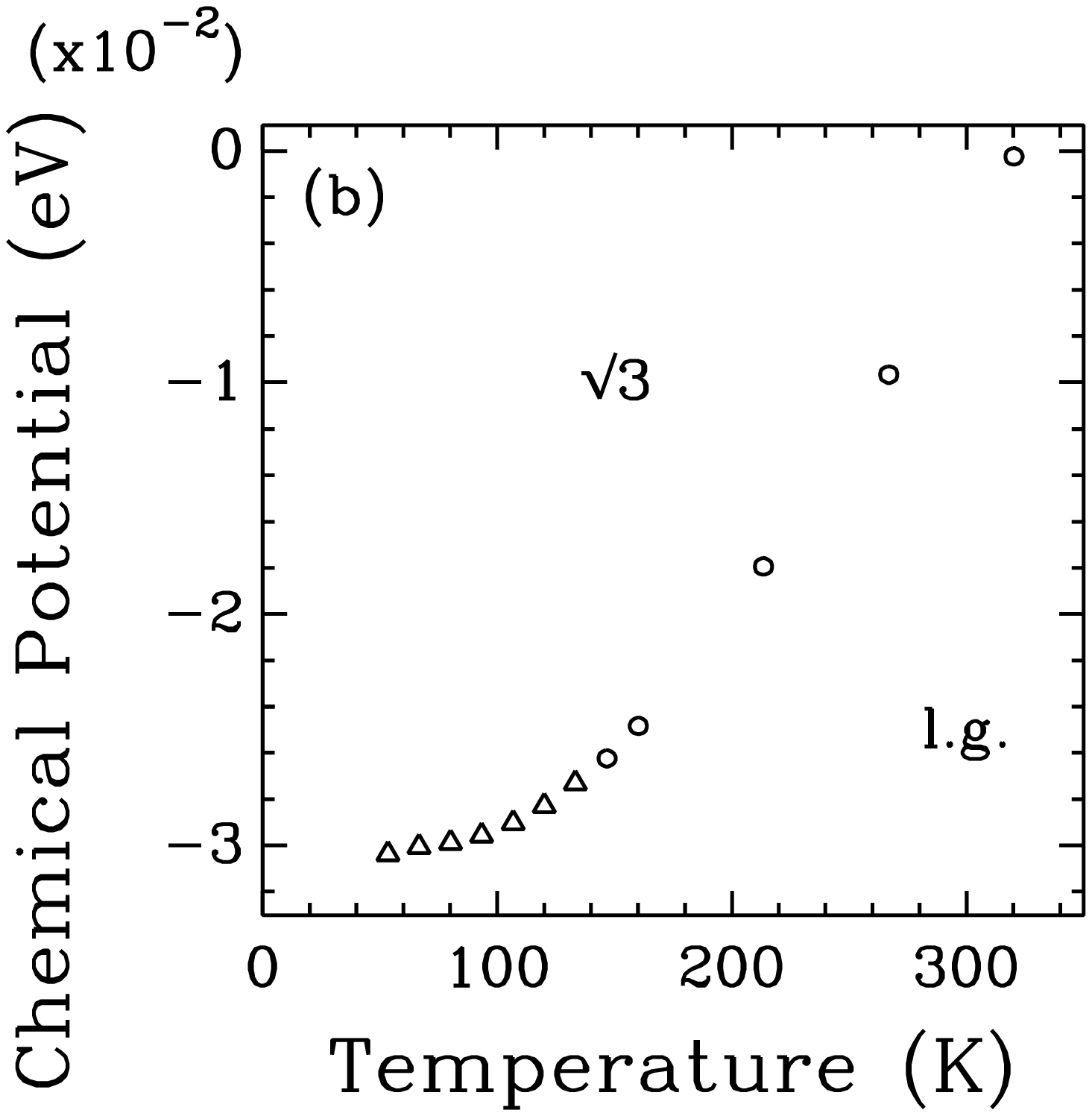}
     \\
    \includegraphics*[bb=48 267 464 695, width = 5.5 cm]{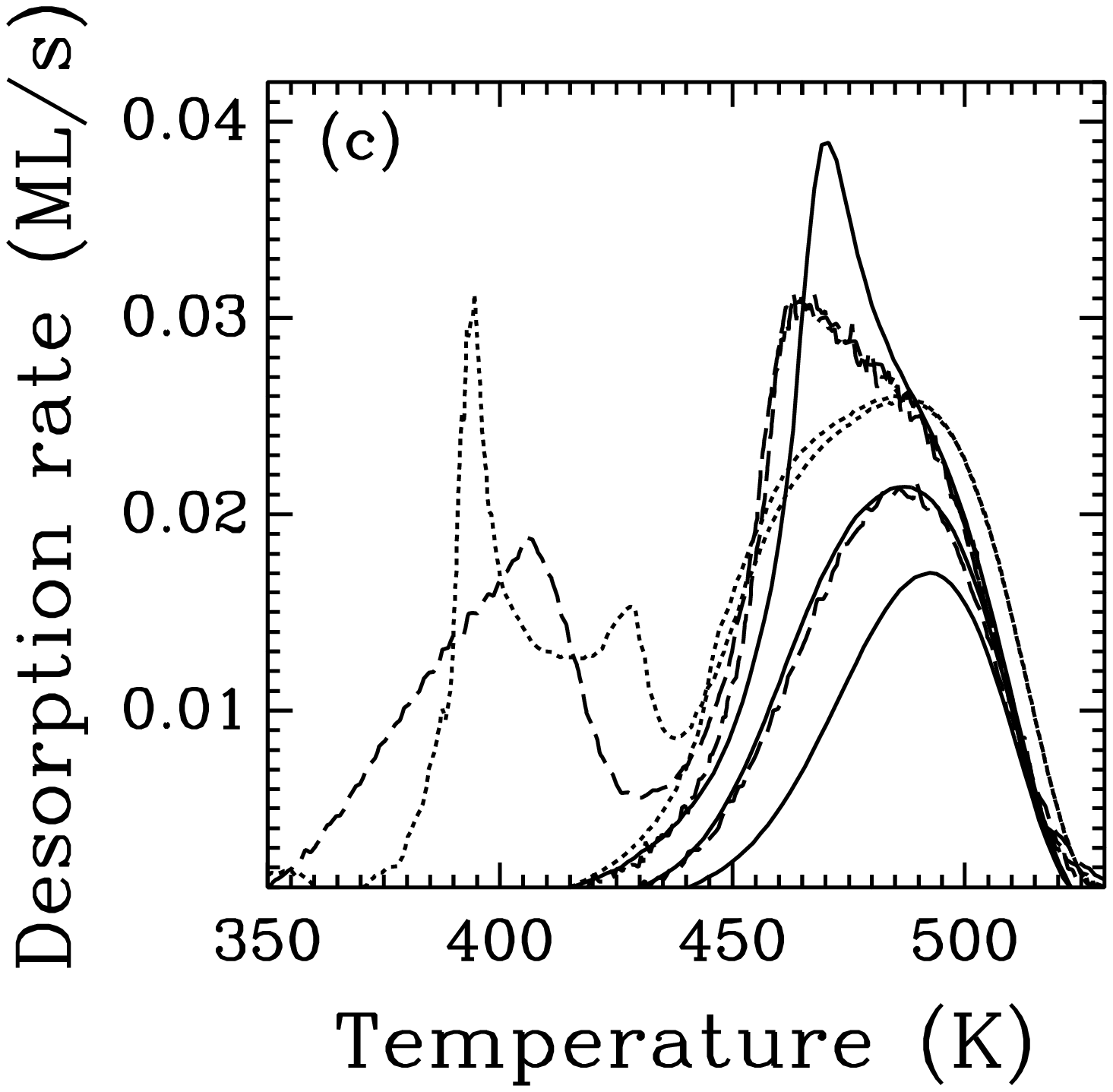}
    \\
    \end{center}
    \caption{(a) Phase Diagram up to 1/3 ML in the temperature-coverage
    plane using the lattice gas model with
    top sites only on a 60$\times$60 lattice (interactions from method A, cf. Table 1).  The
    crosses are the experimental points \cite{pfnur2}, the circles indicate second order
    phase transitions, the triangles first order transitions between lattice gas (l.g.)
    and $(\sqrt{3}\times\sqrt{3})R30^{\circ}$ ($\sqrt{3}$) phases.
    (b) The corresponding phase diagram in the chemical potential-temperature plane.
    (c) The predicted temperature programmed desorption spectra using the experimental
    sticking coefficient \cite{menzelstick} and a heating rate of 5 $K/s$
    at initial coverages of 0.33, 0.25 and 0.15 ML with the top site only model
    (solid), the experimental
    spectra at 0.50, 0.33 and 0.22 ML (dashed) \cite{steinruck2} and the spectra
    with the top and hollow site model at 0.50 and 0.33 ML (dotted).  The theoretical
    curves have had their binding energy adjusted by -174 meV in order to aid the comparison with experiment.}
    \label{coru0001t}
    \end{figure}

     The resulting phase diagram and TPD spectra with the parameters of method A are
     given in Fig.\ref{coru0001t} and show reasonable agreement with experiment provided
     we lower the binding energy of
     a single CO molecule by -174 meV (our unadjusted binding energy places the model curves
     55 K too high).   The experimental sticking coefficient was used \cite{menzelstick}.
     Even with this adjustment there is still some disagreement for the spectra
     with an initial coverage of 1/3 ML.
     One way to correct for this would be to allow for hollow sites in our model for which
     there would be a small but finite number around these temperatures and coverages.

     The phase diagram calculation does not depend on the binding
     energy of the top sites. It is dominated by two phases, the
     $(\sqrt3\times\sqrt3)R30^{\circ}$ structure, which saturates at 1/3 ML and
     a non-ordered lattice gas (l.g.) phase.
     To access the high temperature part of the phase diagram experimentally
     would require pressures up to $10^{-4}$ mbar, which is not accessible
     with LEED \cite{menzel}.  It has in fact been predicted that $T_c$ must
     be at least 550 K at 1/3 ML.  This is in agreement with our calculations, which
     predict an order disorder temperature of 1000 K at 1/3 ML under the restriction
     of on top adsorption only.  We remark that this value will be considerably      lowered
     if there is a spillover into other types of binding sites for coverages beyond
     1/3 ML, resulting in a continuous reduction in temperature
     (and chemical potential) of the hight of the order phase boundary
     around 1/3 ML.  Moreover, the value of $V_2$ can be directly
     read from the experimental phase
     diagram \cite{nagai} with the value of the tricritical point given by
      $\approx 1.06 V_2$.
     Experimentally, the tricritical point is around
     150 K \cite{pfnur2} for which
     there is some evidence from slow diffusion of the CO molecules \cite{menzel}.
     This could be the source of the small discrepancy
     between our calculations and experiment for the first order transition points.

     From these comparisons with experiment we can conclude that DFT does reasonably well in
     at low coverage ($\theta<$ 1/3 ML) and since even in this coverage regime hollow
     sites come into play at high temperatures (T $>$ 150 K)
     it does even better at low temperature.
     Similarly, the adsorption isobars at
     1.3$\times10^{-4}$ mbar and 1.3$\times10^{-6}$ mbar are limited to 1/3 ML for
    T $>$ 400 K.\\ \\
     Moreover, one may question the relevance of incorporating trio interactions
     for coverages
     below 1/3 ML.  Indeed, with such a large nearest neighbor interaction these trios
     will practically never occur below 1/3 ML, even at desorption temperature and despite
     the attractive triangular trio interaction.  However, fitting the calculated
     energies with $V_1$ and $V_2$ alone using method A results in $V_1=0.216$ eV and
     $V_2=-0.060$ eV with $\sigma= 4$ meV.  This would give a phase diagram similar to
     that given in Fig. \ref{coru0001t} but with a tricritical point that one can estimate
     \cite{nagai} to be 75 K instead
     of $\sim$ 120 K as shown in Fig. \ref{coru0001t}.
     For coverages beyond 1/3 ML, it is quite plausible that trio interactions become
     important.  Indeed, this would help to explain the asymmetry in the experimental
     phase diagram.  However, as been noted before, the asymmetry
     of the phase diagram can be explained either with a spillover into other binding sites \cite{piercy}
     or with trio interactions \cite{mcewen},
     but it is quite possible that there is a combination of both.  Since from the
     experimental data we know that a spillover must occur for coverages beyond
     1/3 ML one cannot argue from symmetry arguments alone that one must have trio
     interactions.  This makes it necessary to analyze the experimental data for
     coverages beyond 1/3 ML to consider the relevance of trio interactions for this
     system.  With all of these facts in mind, we now proceed to model the
     adsorption behavior at higher coverages.
    \begin{table}
    \caption{Interactions energies (eV) derived within the DFT framework with three different strategies (A,B,C, cf. text)
    for the CO/Ru(0001) system for top sites together with the corresponding standard deviation ($\sigma$).
    $E_{top}$ denotes the binding energy of the top site,
    the definition of the interaction parameters is given in the text.}
    \begin{tabular}{||c|c|c|c|c|c|c|c|c||} \hline
       Method & $E_{top}$& $V_{1}$ & $V_{2}$ & $V_{3}$ & $V_{lt}$ & $V_{bt}$ & $V_{tt}$ & $\sigma$ (meV)\\\hline
       A & -1.796 & 0.230 & -0.010 & - & - & 0.010 & -0.048 & 3 \\   \hline
       B & -1.948 & 0.137 & 0.002 & 0.022 & 0.104& 0.066 & -0.231 &   2 \\ \hline
       C & -1.948 & 0.263 & 0.002 & 0.022 & -0.021 & 0.004 & -0.042 &   2 \\ \hline
    \hline
    \end{tabular}
    \label{tab;intdft2}
    \end{table}
     \section{Modeling with on-top, hcp and fcc sites}
      The interaction
     energies between sites of the same type for fcc and hcp sites were
     calculated with the same procedure as for the top sites.  For the same reason as
   for the low coverage regime, we present here
     the results of method A only.
     In addition, we show in Table \ref{tab;intdftraw} the adsorption
     energies per CO molecule for all investigated structures.  These
     results are in line with previous DFT calculations that predict that
     the top sites are the most favorable binding site at low coverage
     \cite{norskov,stampfl}.
    \begin{table*}
    \caption{Calculated adsorption energies (eV) for CO/Ru(0001) per CO molecule with method A (cf. text).}
    \begin{tabular}{||c|c|c|c|c|c|c|c|c|c|c||} \hline
       site & $E_a^{\theta=1/9}$& $E_a^{\theta=2/9}$ & $E_a^{\theta=1/4}$ & $E_a^{\theta=1/3}$ & $E_a^{\theta=1/2}$ & $E_a^{\theta=2/3}$ & $E_a^{\theta=3/4}$ &$E_{a,(2\times 2)}^{\theta=1}$&$E_{a, (3\times 3)}^{\theta=1}$\\\hline
       top & -1.796 & -1.686 & -1.793 & -1.827 & -1.572 & -1.452 & -1.371 & -1.172 & -1.176 \\   \hline
       hcp & -1.750  & -1.635 & -1.774 & -1.755 & -1.509& -1.389 & -1.281 & -1.118 & -1.127\\ \hline
       fcc & -1.672 & -1.557 & -1.660 & -1.676 & -1.426 & -1.323 & -1.231 & -1.097 & -1.104 \\ \hline
    \hline
    \end{tabular} \\
    \label{tab;intdftraw}
    \end{table*}
     We note that the energy for the (1$\times$1)-CO
    structure using a $(2\times2)$ and a $(3\times3)$ unit cell differ by 7 meV for
    fcc sites and 9 meV for hcp sites.  When determining the corresponding interactions,
    we needed to include linear trios
    as well as third nearest neighbor interactions to obtain a reasonable value of $\sigma$.
    Thus, we have also included them for the top sites for consistency.  The
    resulting interaction parameters are shown in Table \ref{tab;intdft3}.
         For each interaction
    between different types of sites we used the calculated energy per CO
    molecule ($E_{\rm structure}$)
    for a single co-adsorption configuration in the $p(3\times3)$ cell.
    Each interaction was deduced by:
    $E_{\rm structure}=(E_a+E_b+V_{ab})/2$, where $E_a$ and $E_b$ refer to the adsorption
    energy of the two different binding sites (Table \ref{tab;intdftraw})
    and $V_{ab}$ is the parameter describing the interaction between the two sites
    of different type.
    \begin{table*}
    \caption{On-site ($V_0$) and interactions energies ($V_{1}$,$V_{2}$,$V_{3}$,$V_{lt}$,$V_{bt}$,$V_{tt}$)
    in (eV) to model the CO/Ru(0001) system with top, hcp and fcc sites using method A together with
    the corresponding standard deviation ($\sigma$). In the second part
    interaction energies between sites of different type are compiled. }
    \begin{tabular}{||c|c|c|c|c|c|c|c|c|c||} \hline
       site & $V_0$ & $\Delta E_{\rm site}^{\rm top}$& $V_{1}$ & $V_{2}$ & $V_{3}$ & $V_{lt}$ & $V_{bt}$ & $V_{tt}$ & $\sigma$ (meV)\\\hline
       top & 1.796 & - & 0.220 & -0.010 & 0.001 & 0.008 & 0.014 & -0.061 & 2 \\   \hline
       hcp & 1.750 & 0.046 & 0.240 & -0.001 & -0.008 & 0.034 & 0.006 & -0.102 & 3 \\ \hline
       fcc & 1.672 & 0.124  & 0.225 & -0.001 & 0.004 & 0.015& 0.003 & -0.088 & 3 \\ \hline
    \hline
    \end{tabular} \\
    \begin{tabular}{||c|c|c|c||} \hline
       sites & $V_{1}$ & $V_{2}$ & $V_{3}$ \\\hline
      distance & $\sqrt{3}/3$ & $2\sqrt{3}/3$ & $\sqrt{21}/3$ \\ \hline
       top-hcp & $\infty$ & 0.049 & -0.017\\   \hline
       top-fcc & $\infty$  & 0.046 & -0.014 \\ \hline
       hcp-fcc & $\infty$ & 0.044 & 0.004\\ \hline
        \hline
    \end{tabular}
    \label{tab;intdft3}
    \end{table*}
    \begin{table}
    \caption{Frequencies (cm$^{-1}$) at 1/9 ML coverage obtained within the DFT framework (method A)
    together with experimental results from Refs. \cite{Woll,jakob}.}
   \begin{tabular}{||c|c|c|c|c|c||} \hline
    Site & $\nu_{x}$ & $\nu_{y}$ &$\nu_z$ & $\nu_{vib}$ & $\nu_{rot}$ \\\hline
    top & 46.0 & 40.4 & 392.7 & 1989 & 378.7   \\   \hline
    exp. (1/3 ML) & 46.0 & 46.0 & 445.0 & 2025 & 413.0 \\ \hline
    bridge & 163.0 & 41.8 & 344.3 & 1789 &  170.9    \\   \hline
    fcc &  95.3 & 62.9 & 300.6 & 1759 & 158.3 \\ \hline
    hcp & 149.8 & 140.7 &  313.3 & 1720 & 196.1 \\ \hline
    \end{tabular}
    \label{tab;intfreq}
    \end{table}
     The calculated frequencies at low
    coverage (1/9 ML) are compiled in Table \ref{tab;intfreq}.
    The top site frequencies agree reasonably well with experimental data and
    previous DFT calculations for 1/3 ML \cite{norskov}.
    Until very recently \cite{jacob2}, only one peak
    in the typical
    on-top region was observed by IRAS experiments \cite{menzel2,weinberg2} for coverages exceeding 1/3 ML,
indicating either strong coupling and/or a very low infra red
 intensity of the molecules adsorbed in non atop sites.
On the other hand, hollow and bridge sites
    have been inferred from Helium scattering experiments, but the precise values of their
    frequencies cannot be determined at the present time \cite{Woll}.  However, our
    calculated stretch frequencies for hcp and bridge sites are in good qualitative
    agreement with previous DFT calculations at 1/3 ML that have their stretch
    frequencies at 1800 cm$^{-1}$ and 1885 cm$^{-1}$ respectively \cite{norskov}.


   As can be seen from Table \ref{tab;intdft3}, one has such a small binding difference
     between top and hcp sites that at desorption temperatures (T $>$ 300 K)
     a significant (greater than 0.1 ML) population of hcp sites
     below 1/3 ML occurs, completely at odds with experiment.  The resulting
     isobars from these interaction parameters and frequencies \cite{kreuzer5}
     is shown in Figure 2a, with an adjustment of the binding energy of -174 meV (the same
     shift needed for the TPD spectra).
     At first glance, the curves seem to agree well with the
     experimental data points included in the figure.
     However, a closer investigation reveals, that the inflection
     at 1/3 ML is not captured correctly in the calculation.
    This disagreement is more clearly seen when the resulting
     isosteric heat of adsorption is calculated from these
     isobars using the ASTEK software \cite{astek}: the calculated heat of adsorption
     does not rise at 1/3 ML unlike experiment which rises from $\sim$ 155 kJ/mol
     to over 180 kJ/mol at 1/3 ML as seen in Figure 2b.
     This is a consequence of having a significant population of hcp sites below 1/3 ML.
     It has been argued \cite{kreuzer5} that the experimental value of the heat of adsorption
     is too high by about 8 kJ/mol at 1/3 ML, but this does not improve significantly
     the agreement with our calculations.
     A similar disagreement occurs for the TPD spectra (Figure \ref{coru0001t}c)
     when using the same
     sticking coefficient as with top sites only: The absence of a significant rise
     in the heat of adsorption at 1/3 ML is reflected in
     a shallower minimum in the desorption rate
     around 440 K for initial coverages greater than 1/3 ML.  Correspondingly,
     the calculated rate, at around 450 K, does not peak as sharply as in experiment.
     (The small spike in the desorption rate at 430 K results from
     a small change in the chemical potential at this coverage (0.36 ML).
     An examination of the {\it calculated} sticking coefficient \cite{mcewen2}
     here shows that this spike will be eliminated in a full model calculation.)
     Moreover, the binding difference between fcc and top sites
     is much larger than between hcp and top sites.  This excludes a
     $(2\sqrt{3}\times2\sqrt{3})R30^{\circ}$ structure, for which equal
     population of hcp and fcc sites has to occur at 1/2 ML.

     We also remark
     that the inflexion point at 1/2 ML for both pressure isobars is due
     to a (4$\times$2) structure with equal populations of hcp and top sites.  This
     ordered structure is reflected in the rise of the heat of adsorption
     and its subsequent sharp drop to about 50 kJ/mol around 0.55 ML.
     In the TPD spectra the peak at 390 K confirms
      the presence of the (4$\times$2) structure.
     However, because of fluctuations in the calculated chemical potential
     at this coverage and temperature we have estimated
     the height of the peak to have an uncertainty of 20 \%.  On a more positive note,
     DFT calculations seem to agree with LEED experiments that suggest that
     the hollow sites
     have a binding energy between that of the top sites and the bridge sites,
     the latter been calculated to have a binding difference with respect to the top
     sites of 147 meV (obtained from the adsorption energy per CO molecule
     at 1/9 ML giving $-1.649$ eV). Finally, we note that the
     third nearest neighbor interaction
     between top and hollow sites is attractive which does favor the
     $(2\sqrt{3}\times2\sqrt{3})R30^{\circ}$ structure, but which is not
     sufficiently large to give it.
    \begin{figure} \begin{center}
    \includegraphics*[bb=48 267 464 695, width = 5.5 cm]{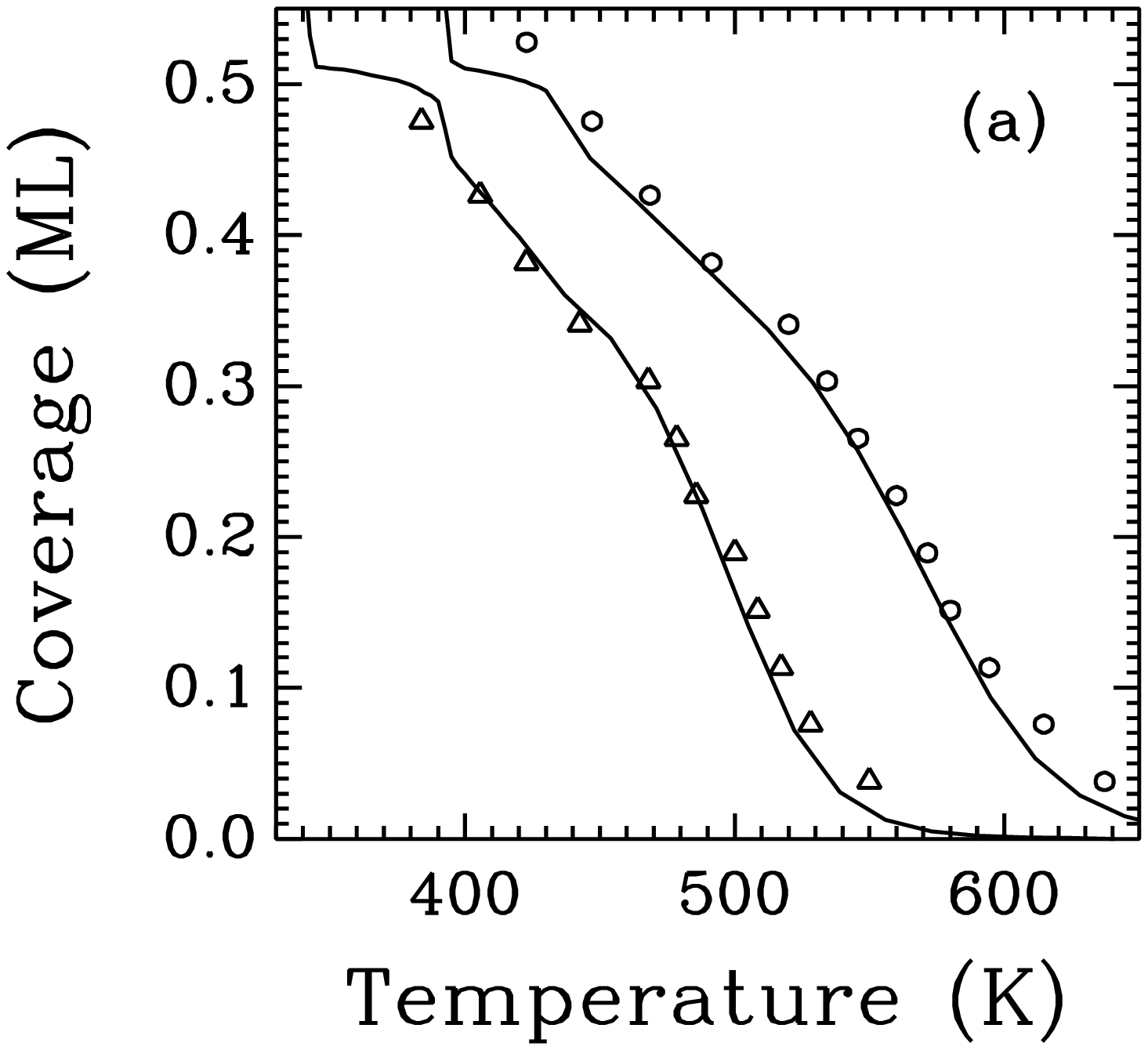}\\
    \includegraphics*[bb=48 267 464 695, width = 5.5 cm]{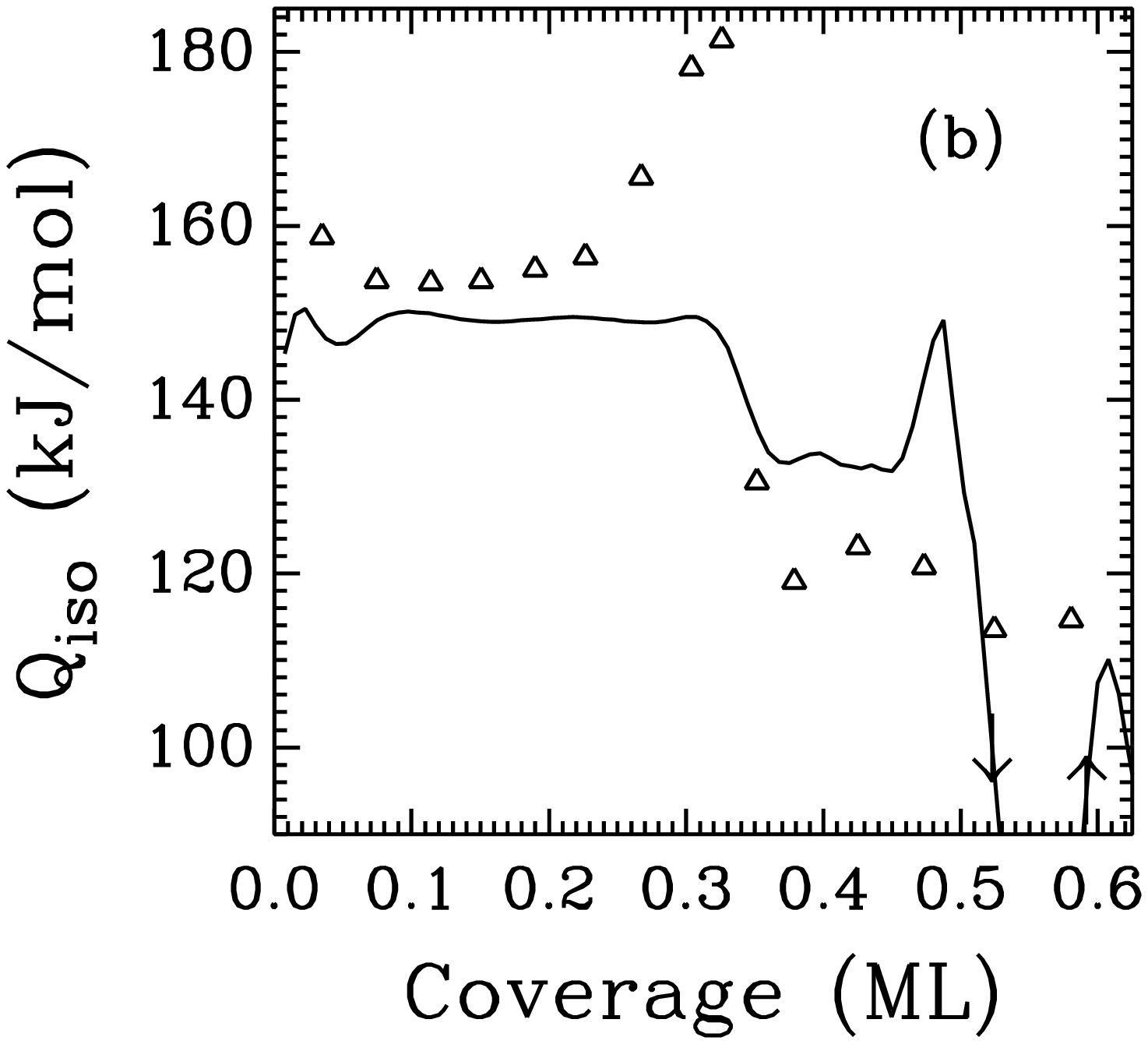}\\
    \end{center}
    \caption{(a) The predicted isobars from a lattice gas model calculation with
    top and hollow sites with interactions calculated from
    DFT (solid curves) compared with experiment at pressures of 1.3$\times10^{-4}$ mbar
    (circles) and 1.3$\times10^{-6}$ mbar (triangles).  The binding energy of the
    isobars has been shifted by -174 meV in order to aid with the comparison with experiment.
    (b) The resulting isosteric heat of adsorption (solid curve)
    from the same model calculated from the isobars and compared with experiment
    (triangles) \cite{menzel3}.}
    \label{coru0001thf}
    \end{figure}
     \section{Conclusions}
     The phase diagram and TPD spectra were calculated from first principles
     and, apart from a difference in overall binding energy, agree well with
     experimental findings up to 1/3 ML if only top sites are included
     in a lattice gas model.
     To model the system beyond 1/3 ML we included top and three fold hollow sites.
     The resulting binding difference between hollow and top sites was determined
     to be too small to reproduce the isobars and heat of adsorption and, because
     of the non-zero binding energy difference between hollow sites, does not reproduce the
     $(2\sqrt{3}\times2\sqrt{3})R30^{\circ}$ as observed in LEED experiments.
     These results seem to indicate that present GGA functionals are able to reproduce
     the interaction between same adsorption sites, but in order to describe the energy
     differences between different adsorption sites further improvement is necessary.
This shortcoming is not only limited to the CO/Ru(0001) system,
but is well known for CO adsorption on transition metals in
general \cite{Feibelmann01,marek}. The reason for this is an insufficient
description of the energy difference between the highest occupied
($2\pi^\ast$) and the lowest occupied molecular orbital
($5\sigma$) of the CO molecule, which contribute with different
intensity at various adsorption sites. Hence, inter-site energy
differences are affected. Recently, first attempts have been
undertaken to cure this problem and the results look very
promising \cite{kresse}.

     \section{Acknowledgments}
      One of the authors (JSM) would like to thank WestGrid of Canada
      for the use of their computer resources as well as
      the Sumner foundation and the National Science and Engineering
      Research Council of Canada for financial support.
    \bibliographystyle{plain}
    
    \end{document}